\documentstyle[12pt]{article}

\begin{document}
\pagestyle{plain}

\newcommand{\be}{\begin{equation}}
\newcommand{\ee}{\end{equation}}
\newcommand{\bea}{\begin{eqnarray}}
\newcommand{\eea}{\end{eqnarray}}
\newcommand{\vp}{\varphi}

\title{Variational Method for Studying Solitons \\ in the
Korteweg-DeVries Equation}
\author{
Fred Cooper \\
{\small \sl Theoretical Division, Los Alamos National Laboratory,}\\
{\small \sl Los Alamos, NM 87545}\\
\and Carlo Lucheroni\\
{\small \sl Dipartimento di Fisica and Sezione, INFN Universita di Perugia,} \\
{\small \sl 06100 Perugia, Italy}\\
\and Harvey Shepard \\
{\small \sl Physics Department, University of New Hampshire,} \\
{\small \sl Durham, NH 03824}\\
\and Pasquale Sodano \\
{\small \sl Dipartimento di Fisica and Sezione, INFN Universita di Perugia,} \\
{\small \sl 06100 Perugia, Italy}}
\maketitle

\begin{abstract}
We use a variational method based on the principle of least action to obtain
approximate time-dependent single soliton solutions to the KdV equation. A
class of
trial variational functions of the form $u(x,t) = - A(t) \exp \left[-\beta (t)
\left|x-q(t)
\right|^{2n} \right]$, with $n$ a continuous real variable, is used to
parametrize time-
dependent solutions. We find that this class of trial functions leads to
soliton-like
solutions for all $n$, moving with fixed shape and constant velocity, and with
energy
and mass conserved. Minimizing the energy of the soliton with respect to the
parameter $n$, we obtain a variational solution that gives an extremely
accurate
approximation to the exact solution.
\end{abstract}

\section{Introduction}
In a previous work \cite{CSLS}, we introduced a post-Gaussian variational
approximation, a continuous family of trial variational functions more general
than
Gaussians, which can still be treated analytically. We demonstrated their
usefulness by
discussing several features of the general nonlinear Schrodinger equation: we
derived
an approximation to the one-dimensional soliton solution and showed the
``universality'' of the critical exponent for blowup in the supercritical case;
for the
critical case we calculated the critical mass necessary for blowup.

In the present work we consider another application of the post-Gaussian
variational
method, deriving an approximation to the one-dimensional KdV soliton. Of course
the soliton solution to this exactly integrable system is well-known and easy
to find by
direct integration. The aim here is rather to provide another illustration of
the power,
generality, and accuracy of our method in a problem with some new features, in
preparation for treating other more difficult problems.

The time-dependent variational principle \cite{Dirac} is very useful for
quantum
mechanics \cite{Jackiw-Kerman} and field theory problems as well as for
studying
nonlinear systems described by a Lagrangian \cite{CLS}. In our recent work we
showed
how one could obtain information about solitons and self-focusing
(blow-up) \cite{CSLS}, \cite{CLS} in the nonlinear Schrodinger equation using a
version of
Dirac's variational principle and a canonical set of variational parameters.
This choice
of parameters led automatically to a conserved energy for the effective
Lagrangian. For
the KdV equation, however, a similar choice of variational trial functions
leads to a
Lagrangian that is not automatically in canonical form. We employ the method of
Faddeev and Jackiw \cite{Jackiw-Faddeev} to explicitly obtain a canonical
Lagrangian
with symplectic structure, which leads to a conserved energy for the
variational
parameters as well as a Hamiltonian structure. Assuming a variational ansatz of
the
form $u(x,t) = - A(t) \exp \left[-\beta (t) \left| x-q(t) \right|^{2n}
\right]$,we find that the
conserved energy of the effective Lagrangian is minimized for $n = 0.877$,
giving an
extremely accurate energy, velocity, and shape for the approximate single
soliton
solution.

\section{Action for the KdV Equation \protect\\ and the Variational Principle}

The starting point for the derivation of the KdV equation is the action

\be
\Gamma = \int L dt, \label{action}
\ee
where $L$ is given by \cite{Drazin}

\be
L = \int \left( \frac{1}{2} \vp_{x} \vp_{t} - (\vp_{x})^{3} - \frac{1}{2}
(\vp_{xx})^{2}
\right) dx. \label{Lagrangian}
\ee
 From this Lagrangian, one can determine the conserved Hamiltonian

\be
H = \int \left[ (\pi \dot{\vp}) - L \right] dx = \int \left[ (\vp_{x})^{3} +
\frac{1}{2}
(\vp_{xx})^{2} \right] dx, \label{Hamiltonian}
\ee
where $\pi \equiv \frac{\delta L}{\delta \vp_{t}}$. Because the action is
stationary
with respect to variation in $\vp$ (i.e. $\delta \Gamma/\delta \vp = 0$), we
obtain
the equation

\be
\vp_{xt} - 6\vp_{x}\vp_{xx} + \vp_{xxxx} = 0,
\ee
which we recognize as the KdV equation with the identification $u\equiv
\vp_{x}$:

\be
u_{t} - 6 u u_{x} + u_{xxx} = 0,
\ee

$\Gamma$ is the action of the system described by $\vp (x,t)$; the variational
principle is a version of Hamilton's least-action principle. In the variational
principle
$\vp$ is an arbitrary square integrable function. To obtain an approximate
solution to
the KdV equation, we consider a restricted class $\vp_{v} (x,t)$, constrained
to a form
that tries to capture the known behavior of the full $\vp$ for the problem at
hand. In
this article we will consider only trial wave functions that are capable of
describing the
motion of an initial configuration which can be reasonably approximated at time
$t=0$
by $u(x,0) = -A \exp\left[-\beta |x-a|^{2n}\right]$ for suitable choice of
$A,\beta,a$
and $n$. Thus, we choose for our trial wave function

\be
u_{v}(x,t) = -A(t) \exp \left[-\beta(t) |x-q(t)|^{2n} \right], \label{u_v}
\ee
where $n$ is an arbitrary continuous, real parameter.

The variational parameters have a simple interpretation in terms of expectation
values with respect to the ``probability''
\be
P(x,t) = \frac{\left[u_{v}(x,t)\right]^{2}}{M(t)},
\ee
where the mass $M$ is defined as

\be
M(t)\equiv \int \left[u_{v}(x,t)\right]^{2} dx.
\ee
(Here we allow $M$ to be a function of $t$, even though $M$ is conserved for
the exact
KdV equation.)

Since $\langle x-q(t) \rangle = 0$, $q(t) = \langle x \rangle$. From (8)
we have

\be
A(t) = \frac{M^{1/2} (2\beta)^{1/4n}}{\left[2\Gamma\left(\frac{1}{2n} +
1\right)\right]^{1/2}}.
\ee
The inverse width $\beta$ is related to

\be
G_{2n} \equiv \langle |x-q(t)|^{2n} \rangle = \frac{1}{4n\beta}.
\ee
Thus, we can write

\be
u = \vp_{x} =  -\frac{M^{1/2} (2\beta)^{1/4n}}{\left[2\Gamma\left(\frac{1}{2n}
+
1\right)\right]^{1/2}} \exp\left[-\beta(t)|x-q(t)|^{2n}\right].
\ee

To evaluate the action we first must determine $\vp$. A convenient choice of
integration constant gives

\bea
\vp(x,t) & = & -A(t) \int_{q(t)}^{x} \exp \left[-\beta(t) |y-q(t)|^{2n}\right]
dy
\nonumber \\
                & = & -A(t) (2n)^{-1} \beta^{-1/2n} \epsilon (x-q(t)) F\left[
\beta^{1/2n} |x-
q(t)| \right],
\eea
where

\be
F\left[ \beta^{1/2n} \left(x-q(t)\right) \right] =
\int_{0}^{\beta|x-q(t)|^{2n}} e^{-z}
z^{\frac{1}{2n} -1} dz.
\ee
The sign function $\epsilon(x)$ explicitly displays the oddness properties of
$\vp$:

\be
\vp\left( x-q(t),t \right) = -\vp \left( -\left[ x-q(t) \right],t \right).
\ee
For $n = 1$ we have

\be
F\left[ \beta^{1/2} \left(x-q(t)\right) \right] = \pi^{1/2} \, {\rm Erf}
\left[\beta^{1/2}
\left(x-q(t)\right) \right] .
\ee
The oddness property of $\vp$ guarantees conservation of $M$ and $\beta$, as we
will see below. (For a two soliton ansatz neither $M$ nor $\beta$ would be
time-
independent.)

We next evaluate the action, given by (\ref{action}) and (\ref{Lagrangian}),
for the
trial wave function in (\ref{u_v}). First consider

\be
\vp_{t} = -\dot{q} u + \left( \frac{\dot{M}}{2M} + \frac{\dot{\beta}}{4n\beta}
\right)
\vp + \dot{\beta} G(x-q(t)), \label{phi_t}
\ee
where

\be
G(x-q(t)) \equiv A \int_{0}^{x-q(t)} |y|^{2n} \exp
\left[-\beta(t)|y|^{2n}\right] dy.
\ee
We notice that $G$ is also odd in $x-q(t)$. Thus, in determining $\int \left(
\frac{1}{2}
\vp_{x} \vp_{t} \right) dx$, only the first term of (\ref{phi_t}) contributes
and we
obtain $-M\dot{q}/2$ for the resulting integral. This oddness property causes
$\beta$
to be a constraint rather than a dynamical variable in the ensuing Lagrangian
dynamics.

Evaluating the other terms in (\ref{Lagrangian}), we obtain

\bea
\Gamma(q,\beta,M,n) & = & \int \left( -\frac{1}{2} M \dot{q} + C_{1}(n)
\beta^{1/4n}
M^{3/2} - C_{2}(n) M \beta^{1/n} \right) dt \nonumber \\
& \equiv & \int L_{1} (q,\dot{q},M,\beta) dt,
\eea
where

\bea
C_{1}(n) & = &\left(\frac{8}{9}\right)^{1/4n} \left[ 2\Gamma\left(\frac{1}{2n}
+
1\right) \right]^{-1/2} \nonumber \\
\\
C_{2}(n) & = & \frac{1}{4} n 2^{1/n} \frac{\Gamma\left(2 -\frac{1}{2n} \right)}
{\Gamma\left(\frac{1}{2n} + 1\right)}. \nonumber
\eea

The variational equations $\delta\Gamma/\delta q_{i} = 0$ yield a set of
equations
for the parameters $q,\beta$ and $M$. However it is more useful to put our
first
Lagrangian $L_1$ into canonical form, so that the conserved Hamiltonian can be
displayed and the {\it unconstrained} Lagrangian can be obtained. Following the
work
of Fadeev and Jackiw \cite{Jackiw-Faddeev}, we first rewrite $L_1$ as

\be
L_{2} = \frac{1}{4} \left(q\dot{M} - \dot{q}M \right) - H_{2}(\beta,M),
\label{L_2}
\ee
where

\be
H_{2}(\beta,M) = -C_{1}(n)\beta^{1/4n} M^{3/2} + C_{2}(n) M \beta^{1/n}.
\ee
We next recognize that $\beta$ is a variable of constraint since $\dot{\beta}$
does not
appear in $L_2$. We eliminate $\beta$ (using $\delta\Gamma/ \delta\beta = 0$)
and
find

\be
\beta  =\left[d(n)\right]^{4n} M^{2n/3},
\ee
where $d(n) = [C_{1}(n)/4C_{2}(n)]^{1/3}$.

We now eliminate $\beta$ in (\ref{L_2}) in favor of $M$ to obtain:

\be
L_{3} = \frac{1}{4} \left(q\dot{M} - \dot{q}M \right) - H_{3}(M),
\label{L_3}
\ee
where

\be
H_{3}(M) = \left( C_{2} d^{4} - C_{1} d\right) M^{5/3}.
\ee

We now have unconstrained Lagrangian dynamics with a conserved Hamiltonian
$H_{3}$, which is the Hamiltonian (\ref{Hamiltonian}) evaluated with the trial
wave
function. The set of Lagrange equations resulting from (\ref{L_3}) are

\bea
\dot{M}  =  0 & \Longrightarrow& M  =  {\rm const.} \nonumber \\
& \Longrightarrow& \beta = {\rm const.}
\eea
and

\be
\dot{q} = \frac{10}{3} M^{2/3} \left[C_{1} d - C_{2} d^{4}\right],
\ee
as well as a conserved energy

\be
E = \left( C_{2} d^{4} - C_{1} d\right) M^{5/3}     \label{E}
\ee
Thus the velocity of the soliton is constant and related to the conserved
energy via

\be
\dot{q} = c = -\frac{10}{3} E M^{-1}. \label{q-dot}
\ee
Thus we have

\bea
u_{v}(x,t)& =& -d(n) M^{2/3} 2^{1/4n}  \left[  2\Gamma\left(1+\frac{1}{2n}
\right) \right]^{-1/2}\times \nonumber \\
&&\mbox{} \exp \left [-d^{4n} M^{2n/3} |x-ct-x_{0}|^{2n}\right],  \label{u_v2}
\eea
with $c$ given by (\ref{E}) and (\ref{q-dot}). To find the ``best soliton'' for
our class of
trial wave functions we need to minimize the energy of the soliton with respect
to
$n$. In Fig.1 we plot the dimensionless energy $E(n)/M^{5/3}$. We see that the
energy
is minimized for $n=0.877$; at that value we find that $E = -0.3925 M^{5/3}$.

The exact single soliton solution of the KdV equation is given by \cite{Drazin}

\be
u(x,t) = -\frac{c}{2} {\rm sech}^{2} \left(\frac{1}{2} c^{1/2} (x-ct-x_{0})
\right),
\label{u(x,t)}
\ee
where $c = (3/2)^{2/3}M^{2/3} = 1.310 M^{2/3}$. Using (\ref{Hamiltonian}), we
find
that the energy of the exact soliton solution is given by

\be
E = -\frac{1}{5}\left(\frac{3}{2}\right)^{5/3}M^{5/3} = -0.3931 M^{5/3}.
\label{exactE}
\ee
Comparing (\ref{E}) and (\ref{exactE}), we find that the energy of the true
soliton is
lower than that of the variational approximation (as it must be), but the error
is only
$0.15\%$. Using the proportionality of $c$ and $E$, we find that the
variational
calculation gives for the soliton velocity at the optimal $n$:

\be
v_{0} = 1.308 M^{2/3}
\ee
which is accurate to $ 0.2\%$.

In Fig. 2 we plot the best variational soliton wave function $u_{v}^{2}$ with
$n=0.877$
and the exact single soliton solution $u^{2}$. To show the accuracy in detail
we plot
$u_{v}^{2}(x) - u^{2}(x)$ in Fig. 3; the worst error (at the origin) is about
$0.5\%$.

 From these calculations we conclude that by optimizing the class of
non-Gaussian
variational vave functions we get an extremely accurate approximate soliton
with
correct velocity and energy.

\section*{Acknowledgements}
This work was supported in part by the DOE and the INFN. C.L. thanks the
``Della
Riccia'' Foundation for partial support and the Santa Fe Institute for its
hospitality.
F.C. thanks the University of Perugia and the Santa Fe Institute for their
hospitality
and Roman Jackiw for crucial discussions. We thank Eliot Shepard for help in
preparing the manuscript.

\end{document}